\documentclass{pasj00}
\draft

\begin{document}
\SetRunningHead{Jiang et al.}{fan-spine dome}

\title{Interaction between an emerging flux region and a pre-existing fan-spine dome observed by \emph{IRIS} and \emph{SDO}}

\author{Fayu \textsc{Jiang},  Jun \textsc{Zhang}, Shuhong \textsc{Yang}}
\affil{Key Laboratory of Solar Activity, National Astronomical Observatories, \\Chinese Academy of Sciences,
20A Datun Road, Chaoyang District, Beijing 100012, China}
\email{jiangfayu@nao.cas.cn, zjun@nao.cas.cn, shuhongyang@nao.cas.cn}


\maketitle

\begin{abstract}

We present multi-wavelength observations of a fan-spine dome in the active region NOAA 11996 with the \textit{Interface Region Imaging Spectrograph} (\emph{IRIS}) and the Atmospheric Imaging Assembly on board the \textit{Solar Dynamics Observatory} (\emph{SDO}) on March 9, 2014. The destruction of the fan-spine topology owing to the interaction between its magnetic fields and an nearby emerging flux region (EFR) is firstly observed.
The line-of-sight magnetograms from the Helioseismic and Magnetic Imager on board the \emph{SDO} reveal that the dome is located on the mixed magnetic fields, with its rim rooted in the redundant positive polarity surrounding the minority parasitic negative fields. The fan surface of the dome consists of a filament
system and recurring jets are observed along its spine. The jet occurring around 13:54 UT is accompanied with a quasi-circular ribbon that brightens in the clockwise direction along the bottom rim of the dome, which may indicate an occurrence of slipping reconnection in the fan-spine topology. The EFR emerges continuously and meets with the magnetic fields of the dome. Magnetic cancellations take place between the emerging negative polarity and the outer positive polarity of the dome's fields, which lead to the rise of the loop connecting the EFR and brightenings related to the dome. A single Gaussian fit to the profiles of the \emph{IRIS} SI IV 1394 \AA\ line is used in the analysis. It appears that there are two rising components along the slit, except for the rise in the line-of-sight direction. The cancellation process repeats again and again. Eventually the fan-spine dome is destroyed and a new connectivity is formed. We suggest that magnetic reconnection between the EFR and the magnetic fields of the fan-spine dome in the process is responsible for the destruction of the dome.

\end{abstract}

\section{Introduction}

It is  known that magnetic reconnection plays an important role in solar activity. Various dynamic phenomena, such as flares, 
soft X-ray jets, have been related to magnetic reconnection occurring in a fan-spine topology \citep{1990ApJ...350..672L}. When a new magnetic flux emerges into a large-scale unipolar magnetic field, the minority polarity becomes parasitic and embedded into the opposite polarity. A dome or separatrix fan surface lies in between the closed and open field lines, with a magnetic null point and a spine field line on the separatrix
surface \citep{1998ApJ...502L.181A, 2009ApJ...704..485T, 2011ApJ...728..103L, 2013ApJ...774..154P}.

Magnetic reconnection occurring in a nullpoint fan-spine topology yields numerous observational emission signatures. Anemone X-ray jets \citep{1994ApJ...431L..51S, 1995Natur.375...42Y} are believed to be strongly associated with the nullpoint reconnection in the corona. Saddle-like loop structures observed by \citet{1999SoPh..185..297F} added a new observational evidence of the nullpoint existence in the solar atmosphere. Circular ribbon flares \citep{2009ApJ...700..559M, 2009ApJ...704..341S, 2012ApJ...760..101W} were suggested to be triggered by the reconnection taking place at the null point of the fan surface. The sequential light-up of the flare ribbon suggested that slipping reconnection \citep{2007Sci...318.1588A, 2014ApJ...791L..13L} happened in the fan quasi-separatrix layers. \citet{2012ApJ...746...19Z} found that the nullpoint reconnection was responsible to the cusp-shaped loop in coronal bright points. A solar explosive event exhibiting the spine-fan topology was reported by \citet{2013ApJ...778..139S}. 
Magnetic field extrapolations in flares and jets had confirmed the presence of a fan-spine topology (e.g., \cite{2000ApJ...540.1126A, 2001ApJ...554..451F, 2008ApJ...673L.211M}) and the topology was widely reproduced by magnetohydrodynamics simulations (e.g., \cite{1996PASJ...48..353Y, 2009ApJ...704..485T, 2013ApJ...771...20M}). The ubiquitous presence of the nullpoint topology \citep{2002SoPh..207..223S} makes the magnetic configuration increasingly used as the initial condition of numerical simulations \citep{2009ApJ...700..559M, 2010ApJ...714.1762P,2013ApJ...774..154P}.

Although the fan-spine magnetic configuration has been studied for many years, previous studies mainly concentrated on the formation of the topology and on the observational features associated with the magnetic configuration. To our knowledge, the observations showing the destruction of a fan-spine topology has not been reported before. In this paper, we present the multi-wavelength observations of a fan-spine dome 
with the \textit{Interface Region Imaging Spectrograph} (\emph{IRIS}; \cite{2014SoPh..289.2733D}) and the \textit{Solar Dynamics Observatory} (\emph{SDO}; \cite{2012SoPh..275....3P}) on March 9, 2014. We investigate the destruction of the fan-spine structure as a result of the interaction between the magnetic fields of the fan-spine dome and a nearby emerging flux region (EFR).

\section{Observations}

The fan-spine dome was observed in the active region NOAA AR 11996 on March 9, 2014 (see Figure~\ref{fig1}). Simultaneous observations taken by the  \emph{IRIS} and the \emph{SDO} are utilized in the analysis. The medium sit-and-stare (fixed-slit) observations  from \emph{IRIS} were made from 11:29 UT to 21:39 UT on March 9, 2014. The field of view (FOV) was $60\arcsec \times 65\arcsec$ with the slit centered at (443\arcsec, 348\arcsec). In this study, we primarily use the calibrated level 2 data of the strong SI IV 1394 \AA\  line and slit-jaw images (SJIs) in the filter of 1400 \AA. The spatial resolution was 0.\arcsec166  per pixel for both the spectra and SJIs. The time cadence of the spectral observations was 16 s, while the 1400 \AA\ SJIs were taken at a cadence of 49 s. Multi-wavelength extreme ultraviolet (EUV) observations in 211 \AA, 193 \AA, 171 \AA\ and 304 \AA\ passbands from the Atmospheric Imaging Assembly (AIA; \cite{2012SoPh..275...17L}) on board the \emph{SDO} are also employed for the analysis. The pixel size and time cadence of AIA images are 0$\arcsec$.6 and 12 s respectively.
The line-of-sight magnetograms are provided by the Helioseismic and Magnetic Imager (HMI; \cite{2012SoPh..275..207S, 2012SoPh..275..229S}) on board the \emph{SDO} with a spatial resolution of 0$\arcsec$.5 per pixel and a cadence of 45 s. 
The coalignment is done by cross-correlation between the \emph{SDO}/AIA 1600 \AA\ passband and \emph{IRIS}/SJIs 1400~\AA\ images.

\section{Results}

\subsection{Multi-wavelength observations of the dome}

\begin{figure}
 \begin{center}
  \includegraphics[viewport=10 29 420 232,width=\textwidth]{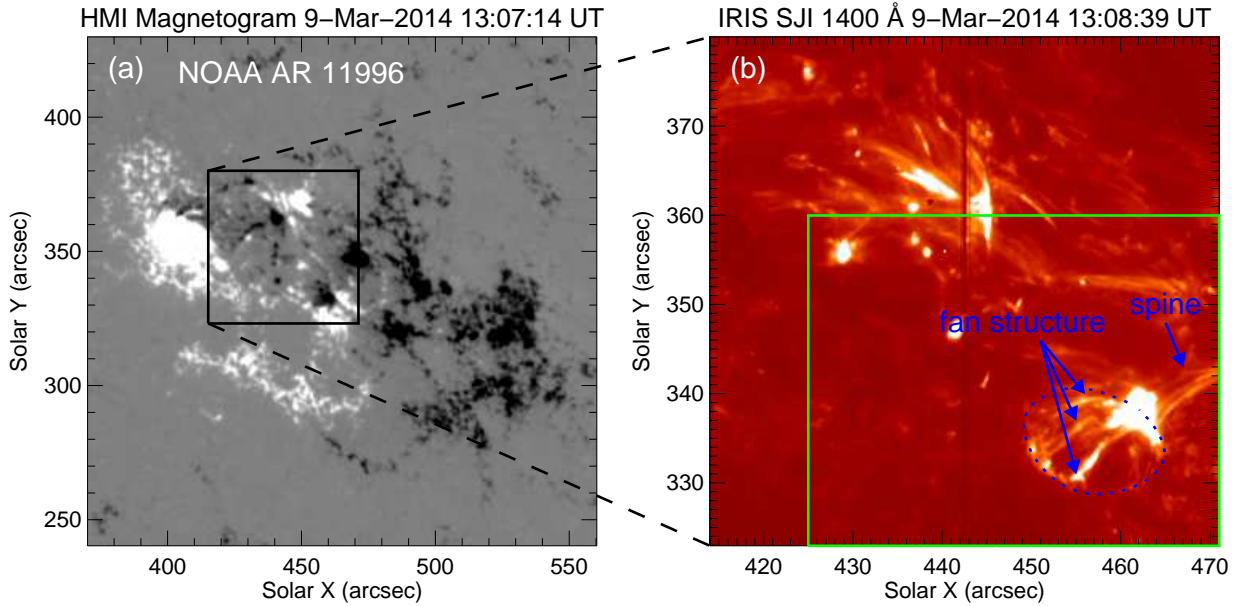} 
 \end{center}
\caption{A magnetogram of AR 11996 from HMI (panel (a)) and a selected \emph{IRIS} 1400 \AA\ SJI (panel (b)) on March 9, 2014. The FOV of the \emph{IRIS} observation is indicated by a black square window in panel (a). The fan-spine dome observed with \emph{IRIS} 1400~\AA\  is shown in panel (b). The spine and fan structure are indicated by blue arrows respectively, while the bottom rim of dome is shown with a blue dotted curve. The green window in panel (b) outlines the region of interest in this study. }\label{fig1}
\end{figure}

Figure~\ref{fig1}(a) displays the overview of AR 11996 with a HMI line-of-sight magnetogram. In this AR, the leading negative magnetic fields are quite diffuse, while there is a main sunspot with positive field in the following polarity.
 The \emph{IRIS} telescope pointed at the interface region of the leading and following polarities in AR 11996, which is shown by a black square window in Figure~\ref{fig1}(a). The fan-spine dome lied in the bottom-right corner of the FOV ($57\arcsec \times 57\arcsec$) of the \emph{IRIS} observation (Figure~\ref{fig1}(b)), which is near the polarity inversion line. 
Frequent emerging fluxes are observed between the fan-spine dome and the main following sunspot. The magnetic evolution in this region is quite complex.
The fan and spine structures can been clearly identified in the image of panel (b) and are indicated by blue arrows respectively. The bottom rim of the dome is plotted with a blue dotted curve. The apparent size of the dome rim was about 16$\arcsec$ in the east-west direction and 12$\arcsec$ in the north-south direction. The region of interest (ROI, $46\arcsec \times 37\arcsec$), which covers the fan-spine dome and a nearby EFR, is labeled by a green window shown in Figure~\ref{fig1}(b).

\begin{figure}
 \begin{center}
  \includegraphics[viewport= 8 8 428 616,width=0.76\textwidth]{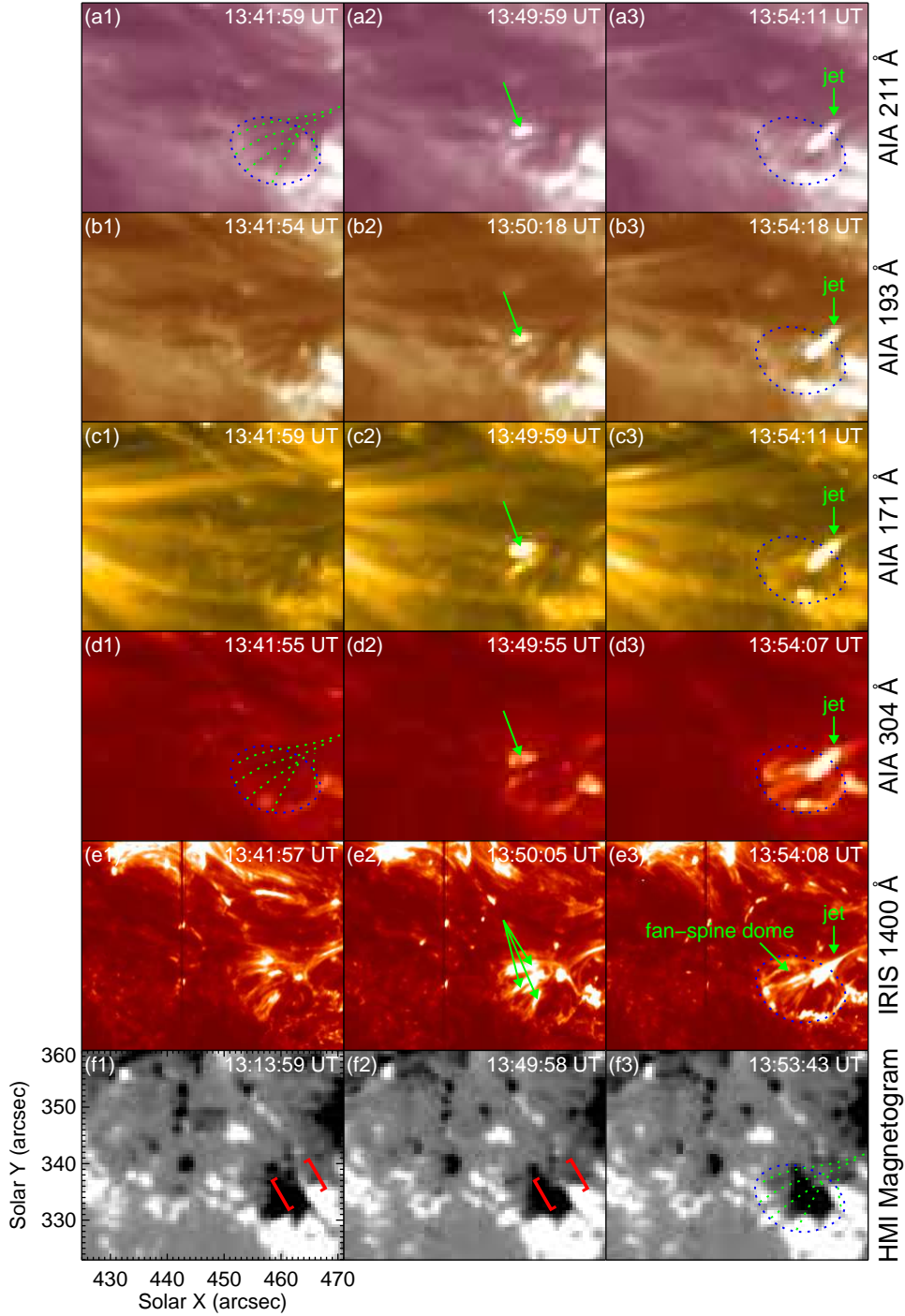} 
 \end{center}
\caption{Multi-wavelength observations of the fan-spine dome and the corresponding magnetic fields with \emph{SDO}/AIA, \emph{IRIS}/SJIs and \emph{SDO}/HMI.
The top four rows show the AIA passband images in 211 \AA, 193~\AA, 171 \AA\ and 304 \AA\ respectively. \emph{IRIS} 1400 \AA\ observations are shown in panels (e1)--(e3). The bottom panels demonstrate the HMI magnetograms. The green curves in panels (a1), (d1) and (f3) outline the framework of the fan-spine structure, while the blue dotted curves show the bottom rim of the dome. The green arrows in the middle column indicate the brightenings of the dome before the occurrence of the jet. An obvious magnetic cancellation site is shown by red brackets ``[ ]" in panels (f1) and (f2).
}\label{fig2}
\end{figure}

Simultaneous multi-wavelength observations combined with \emph{SDO} and \emph{IRIS} are shown in Figure~\ref{fig2}. From top to bottom, the images taken in 211 \AA, 193 \AA, 171 \AA, 304 \AA\ from AIA, in 1400 \AA\ from \emph{IRIS} and the magnetograms from HMI are demonstrated respectively. The characteristic temperatures of AIA emission lines of 211 \AA, 193 \AA, 171 \AA\ and 304 \AA\ are $10^{6.3} K$, $10^{6.2} K$, $10^{5.8} K$ and $10^{4.7} K$, while the formation temperature of SI IV 1400 \AA\ line is $10^{4.8} K$. Some dark filamentary structures, labeled by green dotted curves in Figure~\ref{fig2}(a1) and \ref{fig2}(d1), constitute the framework of the fan-spine dome. These filamentary structures can be seen in all the four AIA channels and in the \emph{IRIS} 1400 \AA\ passband. We overlay the filamentary structures on the magnetograms in panel (f3) and find that every filamentary structure is rooted in the positive magnetic polarity. The bottom rim of the dome is outlined by the bright quasi-circular ribbon shown with the blue dotted curve in panel (e3).
The dome was situated on a mixed magnetic field, with its bottom rim rooted in the positive magnetic polarity surrounding the parasitic negative fields, as shown in panel (f3). Recurring jets \citep{2008A&A...491..279C, 2008A&A...478..907C, 2011RAA....11.1229Y, 2011ApJ...732L...7Y} along the spine were observed around 12:06 UT, 12:16 UT, 12:31 UT, 12:50 UT, and 13:54 UT. The one that took place around 13:54 UT was shown in the third column of Figure~\ref{fig2}. 
At the same time, a quasi-circular ribbon outlined the bottom rim of the dome, as shown in panel (e3).
Before the jet happened, bright features were observed in the fan surface, which are indicated by green arrows in the middle column. 
The opposite magnetic polarities of the dome canceled with each other and an obvious cancellation site was marked with red brackets ``[ ]" in the magnetograms.

\begin{figure}
 \begin{center}
  \includegraphics[viewport= 9 4 415 230,width=0.95\textwidth]{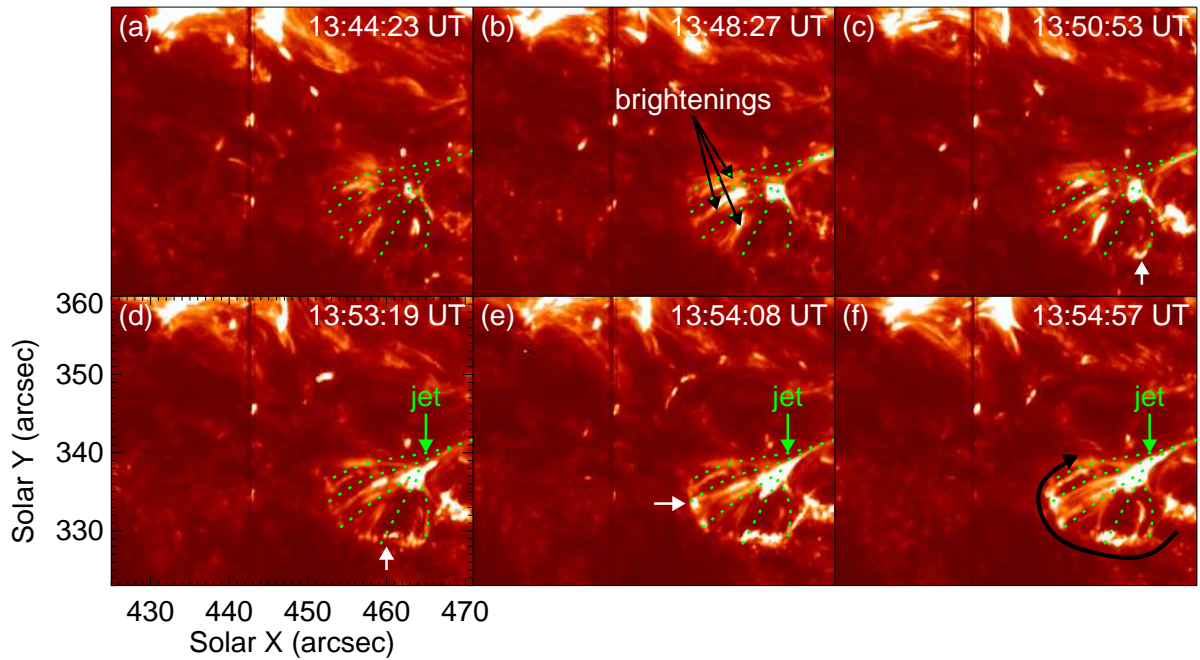} 
 \end{center}
\caption{Observations of a quasi-circular ribbon along the bottom rim of the dome and  a jet along the spine shown with \emph{IRIS} SJIs of 1400 \AA\ passband. The green dotted curves outline the filamentary structures of the dome. The jet is indicated by green arrows in the figure. The quasi-circular ribbon was formed along the clockwise direction, which is shown with the black arrow in panel (f). The white arrows in panels (c)--(e) indicate the brightenings of the filamentary footpoints.
Before the formation of the ribbon and the jet, brightenings turned up between the filamentary structures, which are indicated by the black arrows in panel (b).
}\label{fig3}
\end{figure}

Figure~\ref{fig3} displays more detailed observations of the jet occurring around 13:54 UT and the accompanied quasi-circular ribbon. After a significant brightening at the zenith of the dome at 13:44:23 UT, brightenings appeared between the filamentary structures of the dome four minutes later, as shown in Figure~\ref{fig3}(b). At 13:50:53 UT, a footpoint of the filamentary structures lightened up, as indicated by the white arrow in panel~(c). The jet was initiated from the bright point at the zenith of the dome at 13:53:19 UT, and moved upward along the spine. Meanwhile, the filamentary footpoints brightened in the clockwise direction successively (as indicated by the black arrow in panel (f)), with the bright frontier being denoted by the white arrows in Figures~\ref{fig3}(c)--(e).
A quasi-circular ribbon was formed at the time of 13:54:57 UT. We notice that the ribbon contained several bright kernels, which coincided with the filamentary footpoints of the dome.

\subsection{Emergence of an EFR in the vicinity of the dome}

\begin{figure}
 \begin{center}
  \includegraphics[viewport= 13 15 494 398,width=\textwidth]{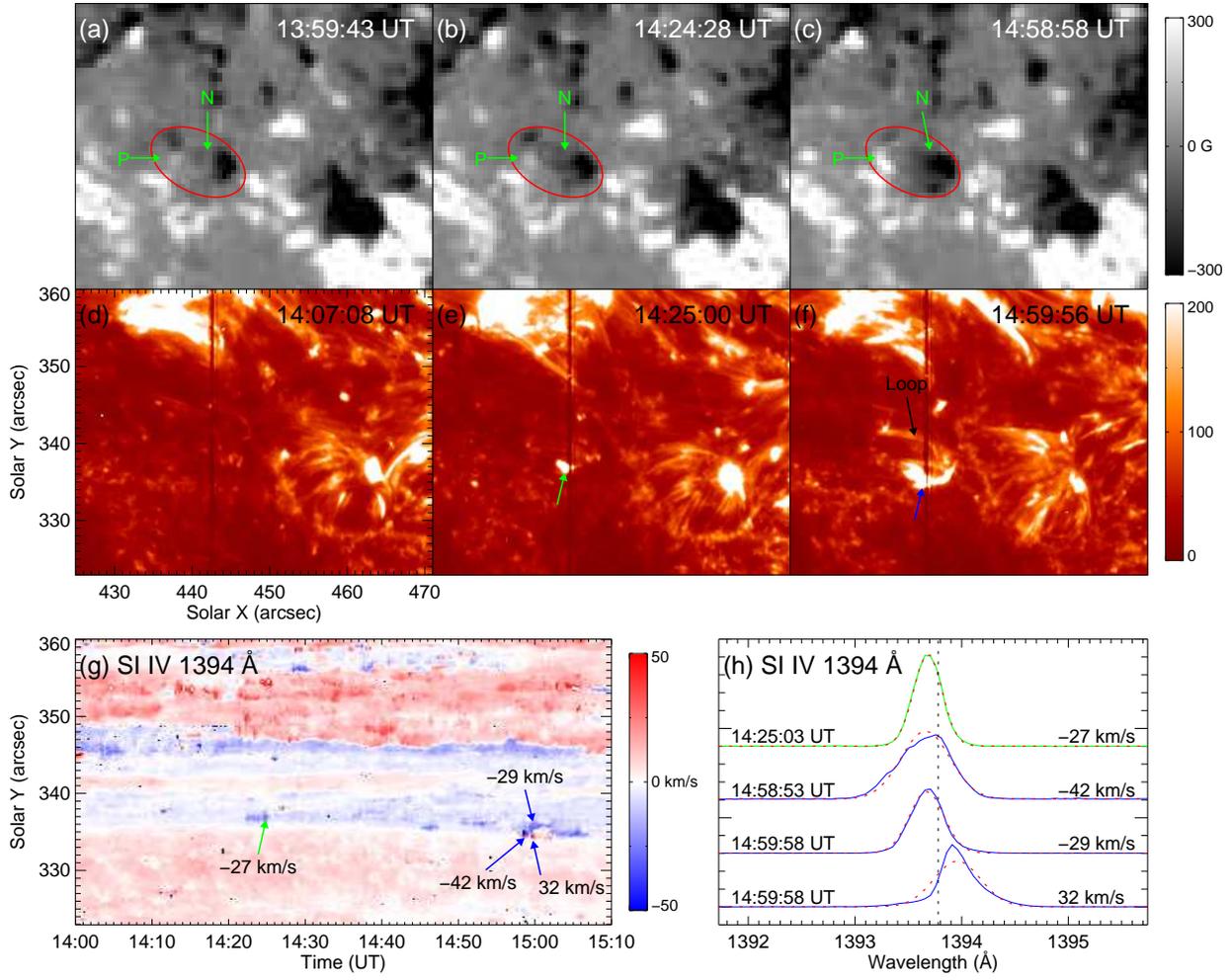} 
 \end{center}
\caption{Panels (a)--(c): emergence of the EFR (labeled by ``P" and ``N") in the vicinity of the dome, which is enclosed by the  red ellipses. Panels (d)--(f) show the IRIS/SJIs 1400 \AA\ observations. The arising loop is indicated by the black arrow in panel (f). The image of SGF Doppler shift of SI IV 1394 \AA\ is plotted in panel (g). The Doppler shift  enhancements, which are related to the appearance of brightenings in panels (e) and (f), are pointed out by green and blue arrows respectively. 
Panel (h) shows the Si IV 1394 \AA\ line profiles at the locations indicated by the four arrows in panel (g). The red dotted lines denote the corresponding SGF results.
}\label{fig4}
\end{figure}

Around 14:00 UT, an EFR emerged near the fan-spine dome. The dipolar patches of the EFR moved apart and  merged with the pre-existing magnetic fields, as  shown in the magnetograms of Figure~\ref{fig4}. We mark the emerging positive and negative polarities as ``P" and ``N" respectively. In this process, a loop arose and became visible, as indicated by the black arrow in panel (f). We apply a single Gaussian fit (SGF) to the SI IV 1394 \AA\ line profiles. The fitted Doppler velocities are shown in panel (g). The green arrows in panels (e) and (g) denote a brightening observed near the footpoint of the loop and a corresponding blueshift enhancement. At 14:59 UT, the blueshift and redshift enhancements (denoted by the blue arrows in panel (g)) in the Dopplergram indicated that there were both upward and downward activities at the bright footpoint of the loop, which is shown by the blue arrow in panel (f). To provide more information at the particular locations marked by the green and blue arrows in panel (g), we display the SI IV 1394 \AA\ line profiles at the four locations in panel (h) respectively. The corresponding SGF results are plotted with the red dotted lines in the panel. The fitted  Doppler velocities are also shown. Comparing the magnetograms (panels (a)--(c)) with the SJIs (panels (d)--(f)), we suggest that these brightenings were induced by the cancellations between the emerging negative flux and the pre-existing opposite polarity.

\begin{figure}
 \begin{center}
  \includegraphics[viewport= 8 7 429 430,width=0.95\textwidth]{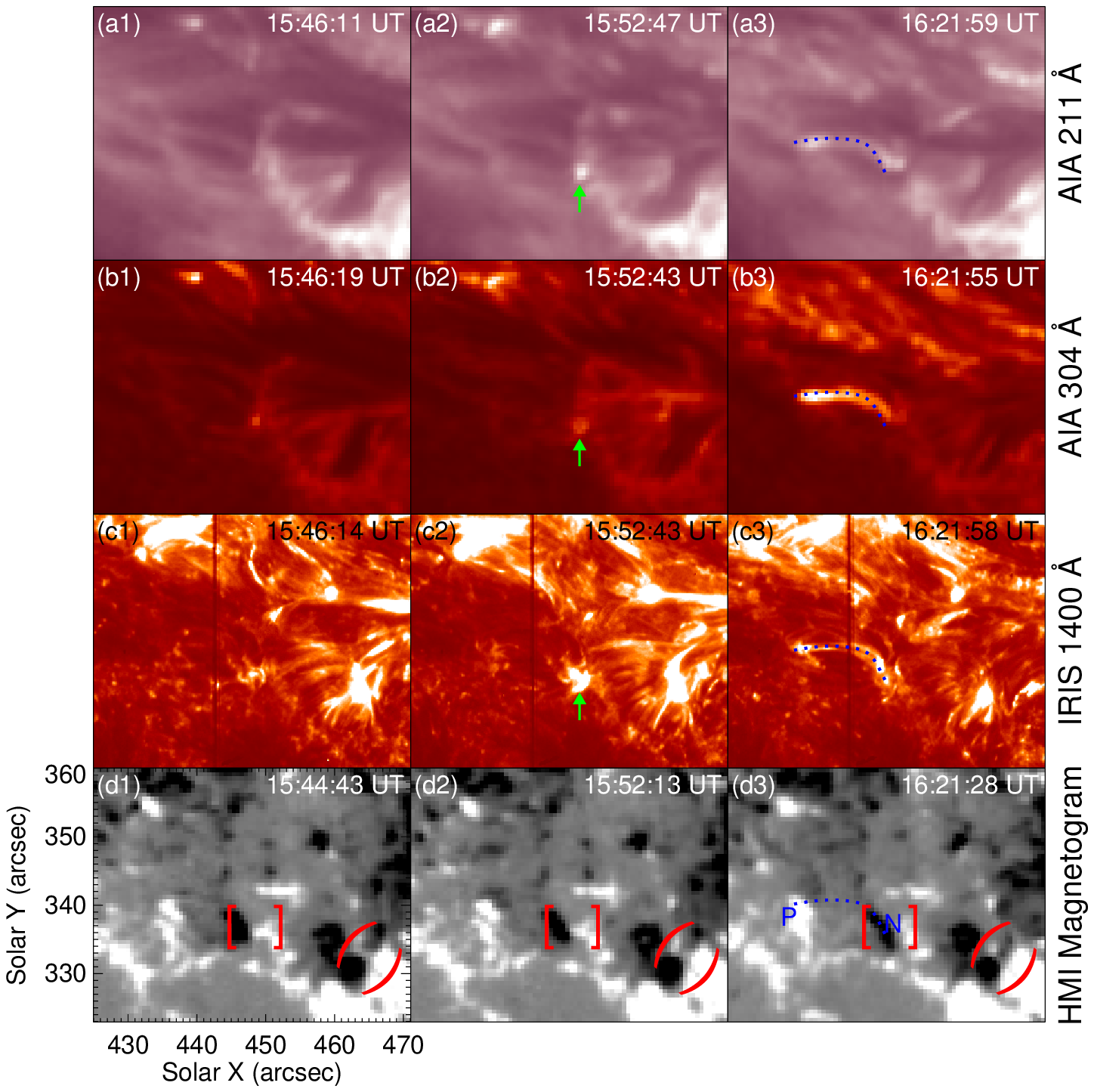} 
 \end{center}
\caption{Magnetic cancellations of the opposite polarities between the EFR and the dome's magnetic fields, together with  the response in the corona and in the chromosphere. The top three rows display the images observed in AIA 211 \AA, 304 \AA, and \emph{IRIS}/SJI 1400~\AA\ respectively. HMI magnetograms are shown in panels (d1)--(d3). The red brackets ``[ ]" indicate the position where the cancellations happen. The brightenings due to magnetic cancellations are indicated by green arrows in panels (a2), (b2), and (c2). The loop brightening is labeled by blue dotted curves in panels (a3), (b3), and (c3) respectively, and the dotted curve is overlaid on the magnetogram (panel (d3)). Magnetic cancellations also take place between the opposite polarities of the pre-existing magnetic fields of the dome (see the red parentheses ``(~)" in panels (d1)--(d3)).
}\label{fig5}
\end{figure}

\subsection{Interaction between the loop connecting the EFR and the dome}

As the dipolar patches of the EFR moved apart, the negative polarity met with the outer positive magnetic fields of the dome. Magnetic cancellations between each other led to brightenings in the location where the loop and the dome contacted, as shown in Figure~\ref{fig5}. The top three rows show the images observed with AIA 211 \AA, 304 \AA, and \emph{IRIS} SJI 1400~\AA\ passbands respectively. The brightenings are indicated by green arrows in the middle column. The loop turned bright around 16:22 UT, which is outlined by the dotted blue curves respectively. We overlay the loop curve on the magnetogram in panel (d3) and find that the loop is rooted at the opposite polarities of the emerging dipole. Magnetic cancellations between the emerging negative polarity and the redundant positive polarity of the dome's fields are illustrated by red brackets ``[ ]" in panels (d1)--(d3). In addition to the cancellations between the emerging flux and the dome magnetic fields, the opposite polarities of the dome also canceled with each other in the meantime. An obvious cancellation site is shown by ``( )" in the figure.

\begin{figure}
 \begin{center}
  \includegraphics[viewport= 18 6 429 330 ,width=0.95\textwidth]{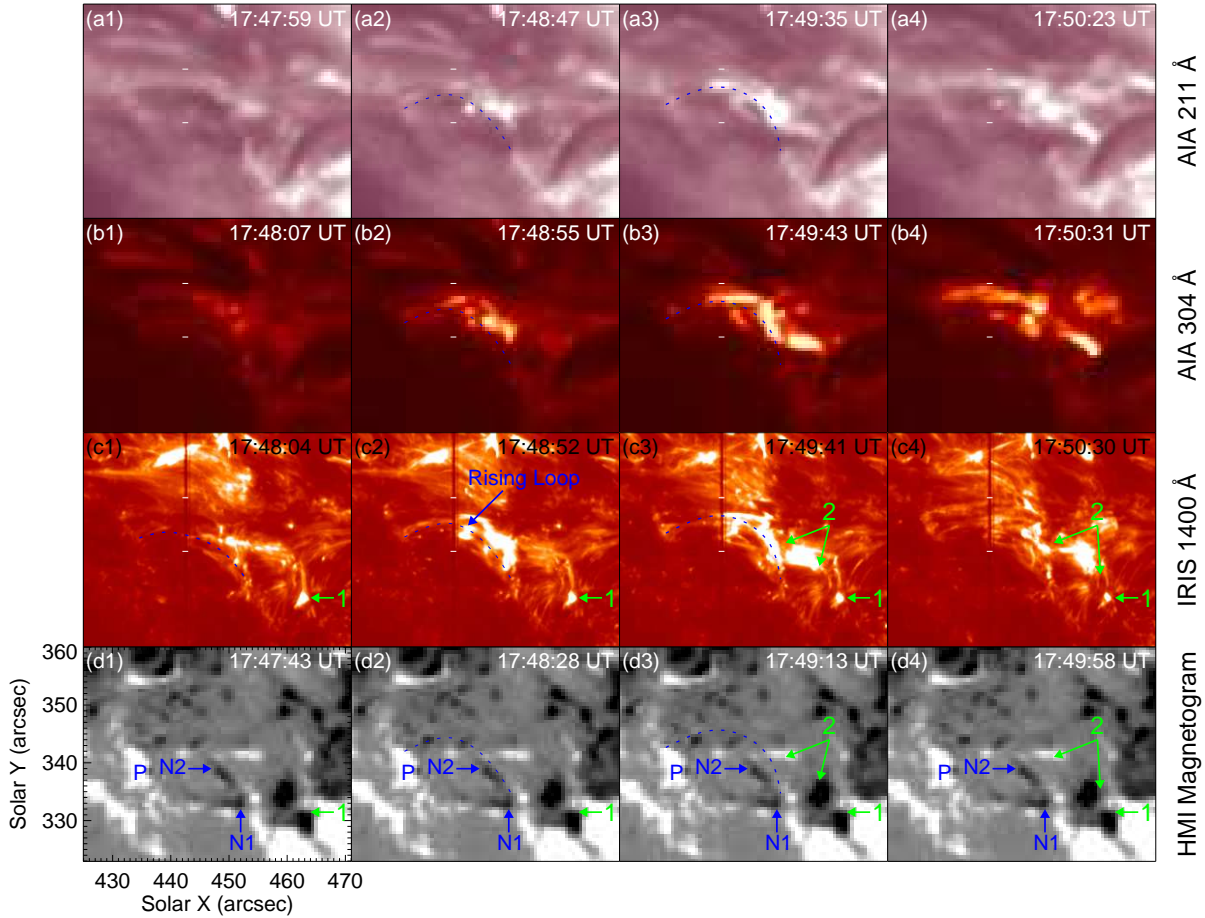} 
 \end{center}
\caption{Simultaneous observations of the rising process of the loop from \emph{SDO}/AIA and \emph{IRIS}, with the corresponding HMI magnetograms. The top three panels show the images from AIA 211~\AA, 304 \AA, and \emph{IRIS}/SJIs 1400 \AA\ passbands respectively. The rising loop is sketched with blue dotted curves in the figure. The two horizontal white bars mark the spatial range of the rising loop. The green arrows labeled with ``1" and ``2" respectively indicate two brightenings related to the dome during this process. The corresponding magnetic fields from HMI are shown in panels (d1)--(d4). The negative polarity ``N" of the emerging dipole has split in two parts ``N1" and ``N2", as shown in the magnetograms. The blue dotted curves and green arrows in the magnetograms are the duplications of those in the SJIs.
}\label{fig6}
\end{figure}

Due to the continuous cancellations between the negative polarity of the EFR and the outer positive magnetic fields of the dome, magnetic energy was released. As a result, the loop began to rise from 17:48 UT to 17:50 UT. Simultaneous observations of AIA 211 \AA, 304 \AA, and \emph{IRIS} SJI 1400 \AA\ are employed to demonstrate the process in Figure~\ref{fig6}. The corresponding magnetograms from HMI are shown in the bottom panels. The emerging negative polarity ``N" had separated into ``N1" and ``N2" before the rise of the loop, as shown in the magnetograms. The rising loop, which is sketched with blue dotted lines in the figure, was rooted in the emerging polarities ``P" and ``N1" (see panels (d2) and (d3)).
The spatial range of the loop along the slit is marked by two horizontal short bars.
The green arrows labeled with ``1" and ``2" indicate two brightenings related to the dome respectively during this process. Brightening ``1" was induced by cancellations between the opposite polarities of the dome's magnetic fields, as indicated by arrow ``1" in the magnetograms. The two ends of brightening ``2" are shown by two arrows respectively. The brightening was moving toward the zenith of the dome during the rising process, which is demonstrated by the inclination of arrows ``2".
The movement of brightening ``2" suggests an energy transfer process.

\begin{figure}
 \begin{center}
  \includegraphics[viewport= 0 15 483 635,width=0.80\textwidth]{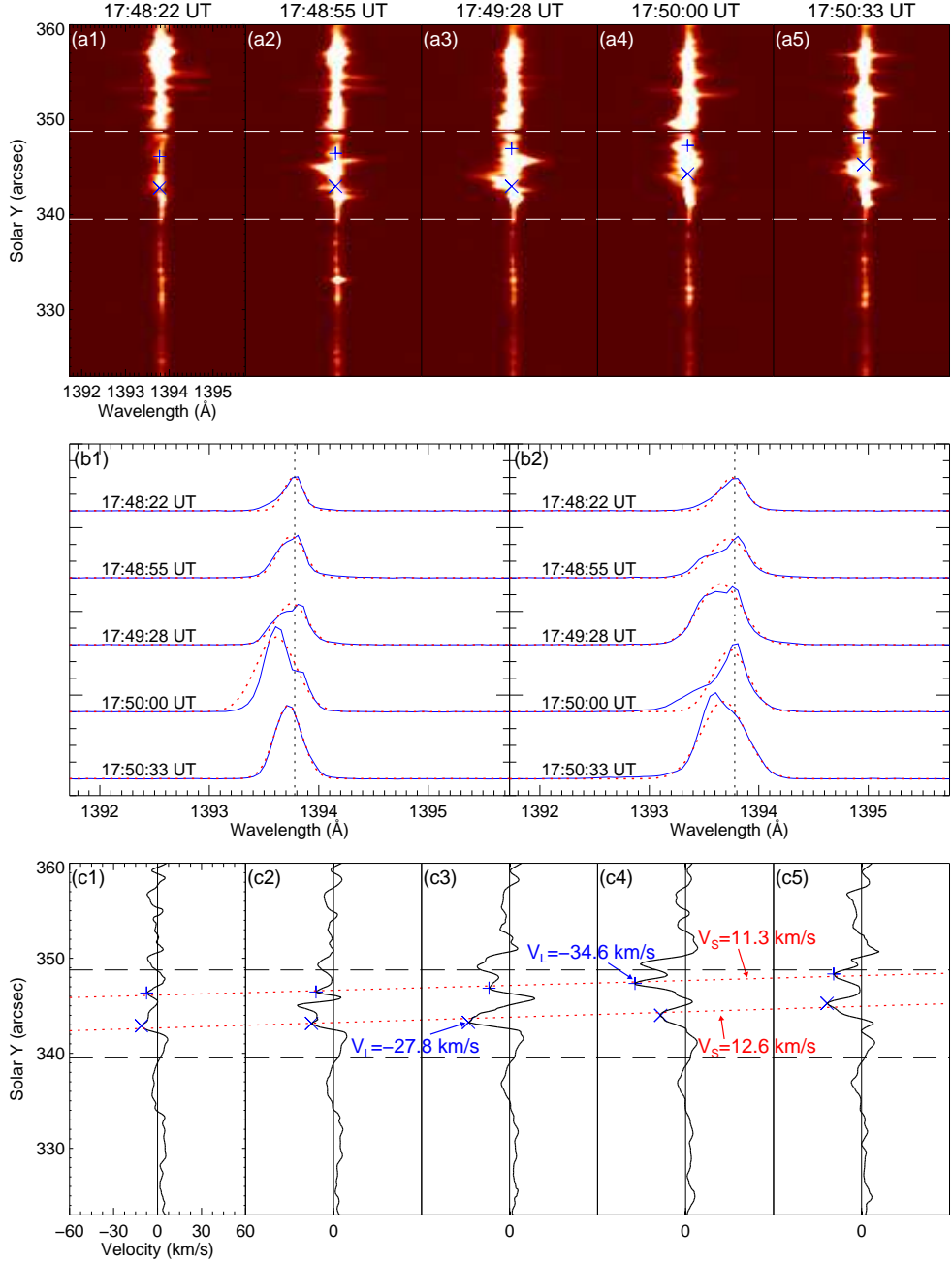} 
 \end{center}
\caption{Panels (a1)--(a5): \emph{IRIS} SI IV 1394 \AA\ spectra associated with the rising loop in Figure~\ref{fig6}. 
The blue lines in panel (b1) represent the line profiles at the five locations marked by the five ``+" symbols in panels (a1)--(a5) respectively, while those in panel (b2) show the profiles at the locations indicated by the five ``$\times$" symbols in panels (a1)--(a5) at each time. The red dotted lines represent the SGF to each profile.
Panels (c1)--(c5): Doppler shift of SI IV 1394 \AA\ at the same time of panels (a1)--(a5). The two horizontal dashed lines are consistent with the two short white bars in Figure~\ref{fig6}. 
The blue ``+" and ``$\times$" symbols mark the local peak blueshifts at each moment. 
The two fitted dotted lines with ascending slope indicate the sub-structures of the loop and the rising components along the slit. The velocities along the slit and the maximum Doppler velocities in the line-of-sight direction are shown in the figure respectively.
}\label{fig7}
\end{figure}

We investigate the spectra of SI IV 1394 \AA\ and the SGF Doppler velocities during the time period. Figures~\ref{fig7}(a1)--(a5) show the SI IV 1394 \AA\ spectra. The spatial range of the two horizontal dashed lines is identical to that of the two short bars in Figure~\ref{fig6}. Significant blueshifts and redshifts turned up in the spectra of the specific spatial range, indicating that both the upward and downward activities took place in the rising process. For each spectrum in panels (a1)--(a5), we select two typical sites, which are marked by blue ``+" and ``$\times$" symbols. The line profiles marked by ``+"  are shown in panel (b1) respectively and the profiles indicated by ``$\times$" are plotted in panel (b2). The dotted red lines represent the corresponding SGF results.
The smoothed SGF Doppler velocities at the same time of the spectra in panels (a1)--(a5) are shown in panels (c1)--(c5). Except for the rise in the line-of-sight direction, which are shown by the blueshifts with ``+" and ``$\times$" symbols, there seems to be two groups of rising components along the slit. So we apply a linear fit with the peak blueshift velocities for each rising component.
The velocities along the slit were 11.3 km s$^{-1}$ and 12.6 km s$^{-1}$ respectively. The maximum rising velocities in the line-of-sight direction were 34.6 km s$^{-1}$ and 27.8 km s$^{-1}$ respectively. The rising velocity includes two portions. One is the line-of-sight velocity, which can be obtained by Doppler shift. The other one is the projection velocity on the solar disk. The ascending velocity along the slit we got in Figure~\ref{fig7} is only one component in the slit directions of the velocity on the solar disk. We infer that the rising loop has fine sub-structures with different velocities. 

\begin{figure}
 \begin{center}
  \includegraphics[viewport= 7 4 550 230,width=\textwidth]{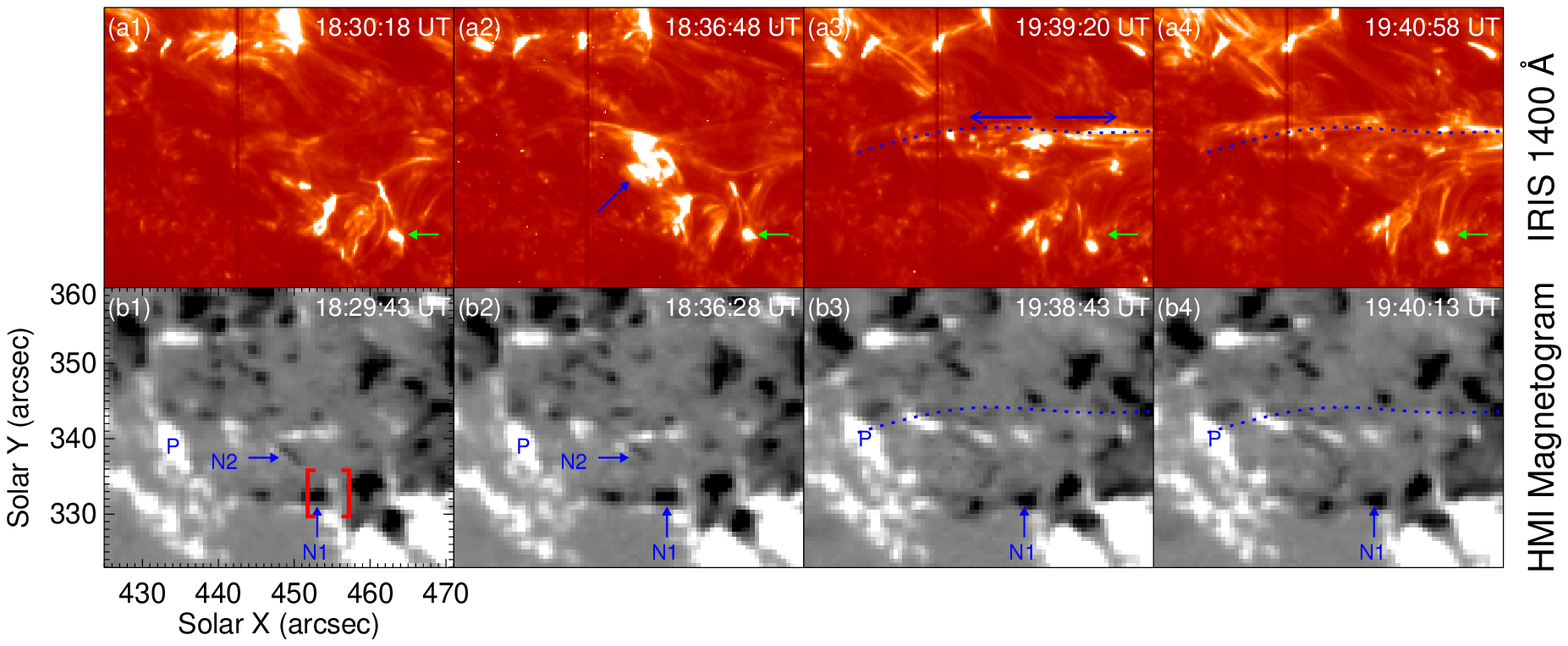} 
 \end{center}
\caption{Panels (a1)--(a4): destruction of the fan-spine dome shown with \emph{IRIS} 1400 \AA\ images. The green arrows show the location of the dome. The blue arrow in panel (a2) indicates a rising process. The blue dotted curves in panels (a3) and (a4) outline the new formed connectivity after the destruction of the dome. Bidirectional flows in the new connectivity are indicated by two blue arrows in panel (a3). Panels (b1)--(b4): corresponding magnetic fields from HMI. Part of the outer positive polarities of the dome have disappeared due to continuous cancellations. The emerging negative polarity ``N1" has merged with the inner part polarity of the dome's magnetic fields (see panels (b3) and (b4)). The blue dotted curves in panels (b3) and (b4) are the duplications of those in panels (a3) and (a4).
}\label{fig8}
\end{figure}

\subsection{Destruction of the fan-spine dome}

The cancellation process described above repeated again and again, as shown in Figures~\ref{fig8}(a1)--(a4). The blue arrow in panel (a2) indicates another loop-rising process due to magnetic cancellations between the EFR and the dome's magnetic fields. Eventually the dome's outer positive polarities on the side of the EFR had partly disappeared due to continuous cancellations. The residual negative flux ``N1" merged with the inner parasitic polarity (see panels (b1)--(b4)). As a result, the fan-spine dome structure was destroyed and a new connectivity was formed. The green arrows in panels (a1)--(a4) indicate the location of the dome. As shown in the SJIs, the fan-spine dome disappeared after 19:39 UT. A new connectivity with one of the footpoints rooted in ``P",  which is outlined by blue dotted lines in panels (a3) and (a4), had been formed since then. Bidirectional flows in the new connectivity were observed and shown by two blue arrows in panel (a3).

\section{Conclusion and discussion}

With the \emph{IRIS} SI IV 1400 \AA\ SJIs and the \emph{SDO}/AIA EUV data, we present multi-wavelength observations of a fan-spine dome in NOAA AR 11996 on March 9, 2014. The destruction of the fan-spine topology is observed for the first time. The bottom of the dome has an apparent size of 16$\arcsec$ and 12$\arcsec$ in the east-west and north-south direction respectively. The dark filamentary structures of the fan surface are identified in the AIA 211~\AA, 193 \AA, 171~\AA, 304 \AA\ channels and in the \emph{IRIS} 1400 \AA\ passband. Each of the filamentary structures is rooted in the outer positive magnetic fields of the dome. The dome lies on a mixed polarity region, with its bottom rim being located on the redundant positive polarity surrounding the minor negative fields. Recurring jets along the spine are observed five times in the observations. The jet around 13:54 UT is accompanied with a quasi-circular ribbon, which brightens in the clockwise direction along the rim of the dome. The ribbon contains several bright kernels, which coincide with the footpoints of the dome's filamentary structures.
Combined with the line-of-sight magnetograms from HMI on board the \emph{SDO}, we study the destruction process of the fan-spine dome induced by magnetic cancellations between the magnetic fields of the dome and the nearby EFR. The chromospheric and coronal response caused by the magnetic interactions are investigated with the multi-wavelength observations and with the SI~IV 1394~\AA\ spectra. As a result of the continuous cancellations between opposite polarities, part of the outer positive polarity of the dome's magnetic fields has disappeared. The fan-spine dome structure is destroyed due to the magnetic reconnection and a new connectivity is formed in the end.

 According to \citet{2009ApJ...700..559M}, in a fan-spine topology with a single nullpoint, a quasi-circular ribbon should be observed in the intersection of the fan with the chromospheric layer when reconnection occurred. 
 The observations show that the fan-spine dome in our study consists of a filament system, which is outlined by the dark filamentary structures in Figure~\ref{fig2}. As the jet is initiated along the spine, sequential brightenings at the footpoints of the filamentary structures in a clockwise direction forms a quasi-circular ribbon. According to the classical flare model, ribbon features are caused by accelerated particles, which flow along the magnetic field lines from the reconnection site, interacting with the solar atmosphere (see \cite{2002A&ARv..10..313P}). Recent studies \citep{2012ApJ...750L...7X, 2014ApJ...788L..18S} revealed that flare ribbons consisted of small-scale bright kernels or knots. The quasi-circular ribbon we observed contains several bright kernels coinciding with the filamentary footpoints. Our observations are in good agreements with both the fan-spine topology and the studies of flare ribbons. We suggest that the dark filaments of the dome outline the open field lines of a fan-spine topology and that the brightenings at the footpoints are caused by accelerated particles along the open field lines due to small-scale magnetic reconnection. The observations that the filamentary footpoints brighten sequentially may indicate a slipping reconnection process \citep{2007Sci...318.1588A, 2009ApJ...700..559M, 2014ApJ...791L..13L}.

Recurring jets initiated from the same location have been widely reported before (e.g., \cite{2008A&A...491..279C, 2008A&A...478..907C, 2011RAA....11.1229Y, 2011ApJ...732L...7Y}) and have been studied by numerical simulations (e.g., \cite{2009A&A...494..329M, 2010A&A...512L...2A, 2010ApJ...714.1762P}).
 It is generally believed that magnetic reconnection is responsible for the  prevalent jets \citep{2007Sci...318.1591S, 2012ApJ...759...33S}. \citet{2010ApJ...714.1762P} performed a three-dimensional numerical simulation with an initial fan-spine magnetic topology. Their model generated recurring jets along open field lines via two distinct regimes of reconnection.
 A quasi-steady magnetic reconnection along the fan surface accumulates the free magnetic energy, then the impulsive reconnection across the null point releases the energy in the form of jet if a threshold amount is reached. We suggest that the recurring jets in this study are triggered by the same mechanism. The repetitive magnetic reconnection near the null point of the fan-spine dome releases the stored magnetic energy and produces intermittent
jets along the open field lines of the fan-spine topology.

In the emergence process of the EFR, an brightening is observed near the footpoint of the emerging loop connecting the EFR, which is shown in Figure~\ref{fig4}(f). Both blueshifts and redshifts of the brightening are observed in the SI IV 1394 \AA\ Doppergram and spectra (see Figures~\ref{fig4}(g) and (h)), which are signatures of upflows and downflows. The upward and downward flows in brightenings are associated with magnetic reconnection, which were widely reported in previous studies (e.g., \cite{2007ApJ...656L..41B, 2010ApJ...723L.169R, 2011ApJ...736...15L}).

The bottom rim of the dome is rooted in the outer redundant polarity, which is in accordance with the fan-spine magnetic topology \citep{1990ApJ...350..672L, 1998ApJ...502L.181A}. The field lines of minority parasitic polarity must come back to the photophere to form closed fields, with a bundle of open field lines immediately outside forming the spine. The destruction of the fan-spine dome structure can be interpreted as follows. The positive open magnetic fields of the fan cancel with the EFR, which makes the open fields weaker and weaker. Preceding cancellations between the EFR and the outer open fields make the incursion into the inner closed 
field lines possible. The loop-rising processes shown in Figure~\ref{fig6} and Figure~\ref{fig8} are induced by magnetic reconnection between the inner closed field lines of the dome and the emerging dipolar field lines. As the confronting positive magnetic fields run out, new reconnection between the EFR and the neighborhood open fields takes place (see Figure~\ref{fig8}(b1)--(b4)). As a result, the dome structure is destroyed and a new connectivity shown in Figure~\ref{fig8}(a3) is formed. 
Various studies (e.g., \cite{1996ApJ...472..840R, 2005GApFD..99...77P, 2007PhPl...14e2106P, 2009PhPl...16l2101P}) showed that the current sheet at the null point led to the collapse of the fan and spine. \citet{2013ApJ...774..154P} performed a resistive magnetohydrodynamical simulation in which a fan-spine topology was disturbed by a boundary-driving velocity. The simulation results showed that a current front transfered along the field lines and focused on the fan surface in the vicinity of the nullpoint. The simulation is similar to the magnetic configuration in this study. We suggest that the formed current sheet at the nullpoint, which is due to magnetic reconnection between the EFR and the dome's magnetic fields, is responsible for the destruction of the fan-spine structure.

\bigskip

This work is funded by the National Natural Science Foundations
of China (11221063, 11203037, 11303050), the
CAS Project KJCX2-EW-T07, the National Basic Research Program of
China under grant 2011CB811403, and the Strategic Priority Research Program -- The Emergence of Cosmological Structures of 
the Chinese Academy of Sciences (No. XDB09000000). The data used in this study is by the courtesy of the \emph{IRIS} teams and the \emph{SDO} teams.

%
%



\begin{thebibliography}{}

\bibitem[Antiochos(1998)]{1998ApJ...502L.181A} Antiochos, S.~K.\ 1998,
\apjl, 502, L181

\bibitem[Archontis et 
al.(2010)]{2010A&A...512L...2A} Archontis, V., Tsinganos, K., \& Gontikakis, C.\ 2010, \aap, 512, L2 

\bibitem[Aulanier et al.(2000)]{2000ApJ...540.1126A} Aulanier, G., DeLuca, 
E.~E., Antiochos, S.~K., McMullen, R.~A., 
\& Golub, L.\ 2000, \apj, 540, 1126 

\bibitem[Aulanier et al.(2007)]{2007Sci...318.1588A} Aulanier, G., Golub, 
L., DeLuca, E.~E., et al.\ 2007, Science, 318, 1588 

\bibitem[Brosius et al.(2007)]{2007ApJ...656L..41B} Brosius, J.~W., Rabin, 
D.~M., \& Thomas, R.~J.\ 2007, \apjl, 656, L41 

\bibitem[Chen et 
al.(2008)]{2008A&A...478..907C} Chen, H.~D., Jiang, Y.~C., \& Ma, S.~L.\ 2008, \aap, 478, 907 

\bibitem[Chifor et 
al.(2008)]{2008A&A...491..279C} Chifor, C., Isobe, H., Mason, H.~E., et al.\ 2008, \aap, 491, 279 

\bibitem[De Pontieu et al.(2014)]{2014SoPh..289.2733D} De Pontieu, B., 
Title, A.~M., Lemen, J.~R., et al.\ 2014, \solphys, 289, 2733 

\bibitem[Filippov(1999)]{1999SoPh..185..297F} Filippov, B.\ 1999, \solphys, 
185, 297 

\bibitem[Fletcher et al.(2001)]{2001ApJ...554..451F} Fletcher, L., Metcalf, 
T.~R., Alexander, D., Brown, D.~S., \& Ryder, L.~A.\ 2001, \apj, 554, 451

\bibitem[Lau 
\& Finn(1990)]{1990ApJ...350..672L} Lau, Y.-T., \& Finn, J.~M.\ 1990, \apj, 350, 672 

\bibitem[Lee et al.(2011)]{2011ApJ...736...15L} Lee, K.-S., Moon, Y.-J., 
Kim, S., et al.\ 2011, \apj, 736, 15 

\bibitem[Lemen et al.(2012)]{2012SoPh..275...17L} Lemen, J.~R., Title, 
A.~M., Akin, D.~J., et al.\ 2012, \solphys, 275, 17

\bibitem[Li 
\& Zhang(2014)]{2014ApJ...791L..13L} Li, T., \& Zhang, J.\ 2014, \apjl, 791, L13

\bibitem[Liu et al.(2011)]{2011ApJ...728..103L} Liu, W., Berger, T.~E.,
Title, A.~M., Tarbell, T.~D., \& Low, B.~C.\ 2011, \apj, 728, 103

\bibitem[Masson et al.(2009)]{2009ApJ...700..559M} Masson, S., Pariat, E., 
Aulanier, G., \& Schrijver, C.~J.\ 2009, \apj, 700, 559 

\bibitem[Moreno-Insertis 
\& Galsgaard(2013)]{2013ApJ...771...20M} Moreno-Insertis, F., \& Galsgaard, K.\ 2013, \apj, 771, 20 

\bibitem[Moreno-Insertis et al.(2008)]{2008ApJ...673L.211M} 
Moreno-Insertis, F., Galsgaard, K., 
\& Ugarte-Urra, I.\ 2008, \apjl, 673, L211 

\bibitem[Murray et 
al.(2009)]{2009A&A...494..329M} Murray, M.~J., van Driel-Gesztelyi, L., \& Baker, D.\ 2009, \aap, 494, 329 

\bibitem[Pariat et al.(2010)]{2010ApJ...714.1762P} Pariat, E., Antiochos, 
S.~K., \& DeVore, C.~R.\ 2010, \apj, 714, 1762 

\bibitem[Pesnell et al.(2012)]{2012SoPh..275....3P} Pesnell, W.~D., 
Thompson, B.~J., \& Chamberlin, P.~C.\ 2012, \solphys, 275, 3

\bibitem[Pontin et al.(2007)]{2007PhPl...14e2106P} Pontin, D.~I., 
Bhattacharjee, A., \& Galsgaard, K.\ 2007, Physics of Plasmas, 14, 052106 

\bibitem[Pontin et al.(2005)]{2005GApFD..99...77P} Pontin, D.~I., Hornig, 
G., 
\& Priest, E.~R.\ 2005, Geophysical and Astrophysical Fluid Dynamics, 99, 77 

\bibitem[Pontin et al.(2013)]{2013ApJ...774..154P} Pontin, D.~I., Priest,
E.~R., \& Galsgaard, K.\ 2013, \apj, 774, 154

\bibitem[Priest 
\& Forbes(2002)]{2002A&ARv..10..313P} Priest, E.~R., \& Forbes, T.~G.\ 2002, \aapr, 10, 313 

\bibitem[Priest 
\& Pontin(2009)]{2009PhPl...16l2101P} Priest, E.~R., \& Pontin, D.~I.\ 2009, Physics of Plasmas, 16, 122101

\bibitem[Rickard 
\& Titov(1996)]{1996ApJ...472..840R} Rickard, G.~J., \& Titov, V.~S.\ 1996, \apj, 472, 840 

\bibitem[Riethm{\"u}ller et al.(2010)]{2010ApJ...723L.169R} 
Riethm{\"u}ller, T.~L., Solanki, S.~K., Mart{\'{\i}}nez Pillet, V., et al.\ 
2010, \apjl, 723, L169 

\bibitem[Scherrer et al.(2012)]{2012SoPh..275..207S} Scherrer, P.~H., 
Schou, J., Bush, R.~I., et al.\ 2012, \solphys, 275, 207

\bibitem[Schou et al.(2012)]{2012SoPh..275..229S} Schou, J., Scherrer, 
P.~H., Bush, R.~I., et al.\ 2012, \solphys, 275, 229

\bibitem[Schrijver 
\& Title(2002)]{2002SoPh..207..223S} Schrijver, C.~J., \& Title, A.~M.\ 2002, \solphys, 207, 223

\bibitem[Sharykin 
\& Kosovichev(2014)]{2014ApJ...788L..18S} Sharykin, I.~N., \& Kosovichev, A.~G.\ 2014, \apjl, 788, L18

\bibitem[Shibata et al.(1994)]{1994ApJ...431L..51S} Shibata, K., Nitta, N., 
Strong, K.~T., et al.\ 1994, \apjl, 431, L51 

\bibitem[Shibata et al.(2007)]{2007Sci...318.1591S} Shibata, K., Nakamura, 
T., Matsumoto, T., et al.\ 2007, Science, 318, 1591 

\bibitem[Singh et al.(2012)]{2012ApJ...759...33S} Singh, K.~A.~P., Isobe, 
H., Nishizuka, N., Nishida, K., \& Shibata, K.\ 2012, \apj, 759, 33 

\bibitem[Su et al.(2009)]{2009ApJ...704..341S} Su, Y., van Ballegooijen, 
A., Schmieder, B., et al.\ 2009, \apj, 704, 341

\bibitem[Sun et al.(2013)]{2013ApJ...778..139S} Sun, X., Hoeksema, J.~T., 
Liu, Y., et al.\ 2013, \apj, 778, 139 

\bibitem[T{\"o}r{\"o}k et al.(2009)]{2009ApJ...704..485T} T{\"o}r{\"o}k,
T., Aulanier, G., Schmieder, B., Reeves, K.~K.,
\& Golub, L.\ 2009, \apj, 704, 485

\bibitem[Wang 
\& Liu(2012)]{2012ApJ...760..101W} Wang, H., \& Liu, C.\ 2012, \apj, 760, 101

\bibitem[Xu et al.(2012)]{2012ApJ...750L...7X} Xu, Y., Cao, W., Jing, J., 
\& Wang, H.\ 2012, \apjl, 750, L7

\bibitem[Yang et al.(2011a)]{2011RAA....11.1229Y} Yang, L.-H., Jiang, Y.-C., 
Yang, J.-Y., et al.\ 2011a, Research in Astronomy and Astrophysics, 11, 1229 

\bibitem[Yang et al.(2011b)]{2011ApJ...732L...7Y} Yang, S., Zhang, J., Li, 
T., \& Liu, Y.\ 2011b, \apjl, 732, L7

\bibitem[Yokoyama 
\& Shibata(1995)]{1995Natur.375...42Y} Yokoyama, T., \& Shibata, K.\ 1995, \nat, 375, 42 

\bibitem[Yokoyama 
\& Shibata(1996)]{1996PASJ...48..353Y} Yokoyama, T., \& Shibata, K.\ 1996, \pasj, 48, 353 

\bibitem[Zhang et al.(2012)]{2012ApJ...746...19Z} Zhang, Q.~M., Chen, 
P.~F., Guo, Y., Fang, C., \& Ding, M.~D.\ 2012, \apj, 746, 19 

\end{thebibliography}
\end{document}